\pdfoutput=1 
\documentclass[12pt]{article}

\usepackage{cite}
\usepackage{url }
\usepackage{graphicx}

\newcommand\ul{\mu_{\rm UL}}
\newcommand\xcrit{x_{\rm crit}}
\newcommand\const{c}
\newcommand\cls{CL$_{\rm S}$}

\begin{document}

\title{\bf \Large
Negatively Biased Relevant Subsets Induced by the Most-Powerful
One-Sided Upper Confidence
Limits \\ for a Bounded Physical Parameter}

\author{Robert D. Cousins\thanks{cousins@physics.ucla.edu}
\\ Department of Physics and Astronomy\\ University of California,
Los Angeles, California 90095, USA}
\date{September 9, 2011}
\maketitle

\begin{abstract}
Suppose an observable $x$ is the measured value (negative or
non-negative) of a ``true mean'' $\mu$ (physically {\em non}-negative)
in an experiment with a Gaussian resolution function with known fixed
rms deviation $\sigma$. The most powerful one-sided upper confidence
limit at 95\% confidence level (C.L.) is $\ul = x+1.64\sigma$, which I
refer to as the ``original diagonal line''.  Perceived problems in HEP
with small or non-physical upper limits for $x<0$ historically
led, for example, to substitution of $\max(0,x)$ for $x$, and
eventually to abandonment in the Particle Data Group's Review of
Particle Physics of this diagonal line relationship between $\ul$
and $x$.  Recently Cowan, Cranmer, Gross, and Vitells (CCGV) have
advocated a concept of ``power constraint'' that when applied to this
problem yields variants of diagonal line, including $\ul = \max(-1,x)
+ 1.64\sigma$.  Thus it is timely to consider again what is
problematic about the original diagonal line, and whether or not
modifications cure these defects.  In a 2002 Comment, statistician Leon Jay Gleser
pointed to the literature on {\em recognizable} and {\em relevant
subsets}. For upper limits given by the original diagonal line, the
sample space for $x$ has recognizable relevant subsets in which the
quoted 95\% C.L. is {\em known} to be negatively biased
(anti-conservative) by a finite amount for {\it all} values of $\mu$.
This issue is at the heart of a dispute between Jerzy Neyman and Sir
Ronald Fisher over fifty years ago, the crux of which is the relevance
of pre-data coverage probabilities when making post-data inferences.
The literature describes illuminating connections to Bayesian
statistics as well.  Methods such as that advocated by CCGV have 100\%
unconditional coverage for certain values of $\mu$ and hence
formally evade the traditional criteria for negatively biased relevant
subsets; I argue that concerns remain.  Comparison with frequentist
intervals advocated by Feldman and Cousins also sheds light on the
issues.
\end{abstract}

\maketitle
\clearpage

\section{Introduction}
\label{intro}

In high energy physics (HEP), a prototype problem with far-reaching
implications and generalizations is that in which an observable $x$ is
the measured value (negative or non-negative) of a ``true mean'' $\mu$
(physically {\em non}-negative) in an experiment with a Gaussian
resolution function with fixed rms deviation $\sigma$, assumed known
for most of this discussion.  Typically the scientific context has
been searches to establish a non-zero value of $\mu$ that would signal
a discovery (non-zero neutrino mass; existence of a rare process;
etc.).  In the absence of a signal, traditionally one would set an
upper limit $\ul$ on $\mu$ at specified confidence level (C.L.),
\begin{equation}
\label{ul95}
\ul = x+1.64\sigma \ {\rm(95\% C.L.),}  
\end{equation}
or $\ul = x+1.28\sigma$ (90\% C.L.). I refer to this method as the
``original diagonal line'', defined by one-tailed integrals with 5\%
and 10\% tail probabilities, respectively.

\begin{figure}
\centering
  \includegraphics*[width=1.0\textwidth,angle=90]{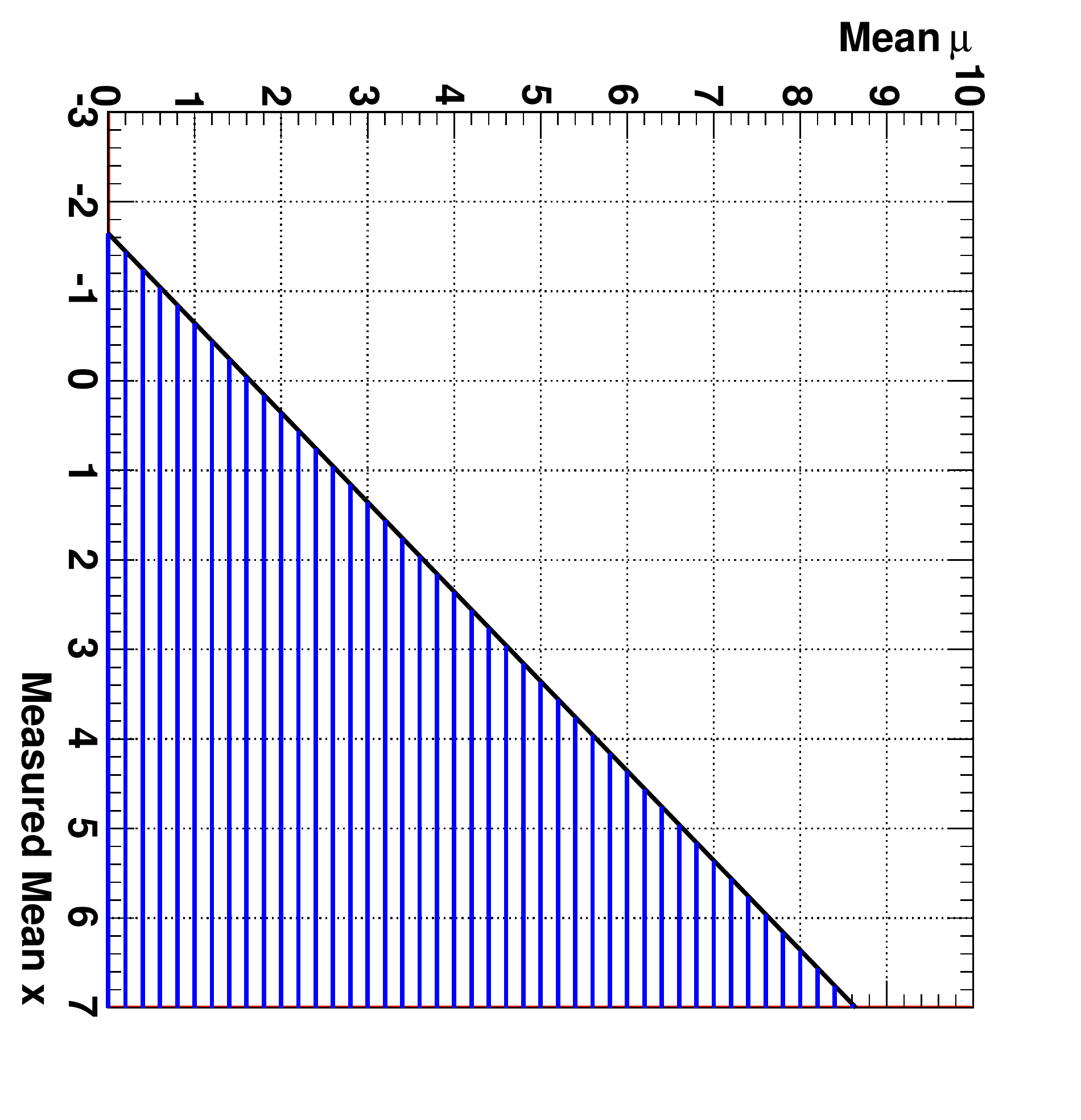}
\caption{Confidence belt corresponding to the original diagonal line
at 95\% C.L., as described in the text. Negative values of $\mu$ do
not exist in the model, so for $x<-1.64$, the set of values of $\mu$
not excluded is the empty set.  Here and in other figures, $\sigma=1$ without
loss of generality. (Equivalently, $\mu$ and $x$ are $\mu/\sigma$ and
$x/\sigma$, respectively.)}
\label{fig-uppershort}
\end{figure}

Figure~\ref{fig-uppershort} displays Eqn.~\ref{ul95} in the form of a
{\em confidence belt}.  (In the figure and much of this paper,
$\sigma$ is set to 1 without loss of generality. Equivalently, $\mu$
and $x$ are to be interpreted as $\mu/\sigma$ and $x/\sigma$,
respectively.)  For each possible value of the unknown true value of
$\mu$ (vertical axis), there is a horizontal line (``acceptance
interval'', drawn for representative values of $\mu$) such that there
is a 95\% probability that the observed $x$ is within that line.  Upon
observing a value of $x$, one draws a vertical line through the
observed value.  The quoted confidence interval for $\mu$ consists of
those values of $\mu$ for which the associated horizontal line is
intersected by the vertical line, in this case thus recovering
Eqn.~\ref{ul95}.  For $x<-1.64$, the confidence interval is thus the
empty set.  Nonetheless, this confidence belt has the property that no
matter what the true value of $\mu$ is, 95\% of the quoted confidence
intervals will contain (``cover'') that value.  Furthermore, this belt
gives the tightest limits (corresponding to a ``most powerful'' test)
of all one-sided belts.

Historically, as $x$ became small, negative, or very negative,
increasing levels of discomfort would set in among many physicists.
When the formal results from using Eqn.~\ref{ul95} yielded $\ul<0$,
some described the upper limit as ``unphysical'' rather than the empty
set, but in any case the experimenter was faced with a problem.  In a
1986 note \cite{VH86}, Virgil Highland summarized six recipes (here
converted to 95\% C.L. if used), three based on the diagonal line of
Eqn.~\ref{ul95}.

\begin{figure}
\centering
  \includegraphics*[width=1.0\textwidth,angle=90]{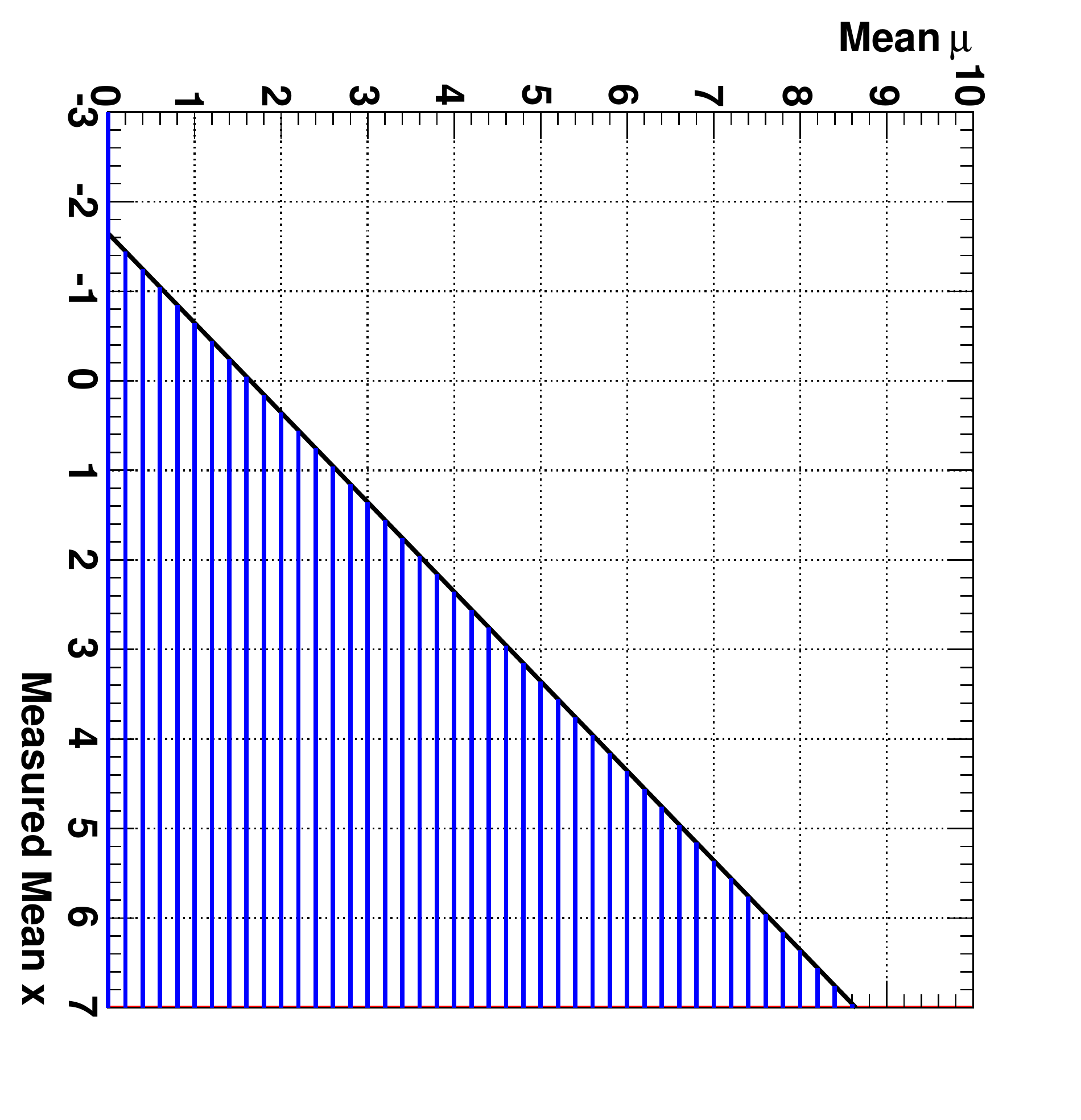}
\caption{Modification of the confidence belt in
Fig.~\ref{fig-uppershort} by replacing the empty-set interval for
$x<-1.64$ with the single point $\ul=0$.  The acceptance interval for
$\mu=0$ thus contains 100\% of the probability for $x$ values rather
than 95\%.  In 1986, Highland \cite{VH86} referred to this as the
``Truncated Classical Method''.  The same result comes from the
derivation in the 2008--2009 ATLAS studies \cite{Aad:2009wy}.}
\label{fig-trunc}
\end{figure}

One possibility, referred to by Highland as the ``Truncated Classical
Method'', was to replace the negative or empty-set upper limit of
Eqn.~\ref{ul95} (obtained when $x<-1.64$) with $\ul = 0$.  That is,
$\ul = \max(0,x+1.64)$, with the corresponding confidence belt shown
in Fig.~\ref{fig-trunc}.  I do not know if $\ul=0$ was ever used in
a publication.  In the 2008--2009 Higgs statistical combination study
in Ref.~\cite{Aad:2009wy}, ATLAS describes a method which again yields
$\ul = \max(0,x+1.64)$.  This method is used by the 2011 ATLAS
supersymmetry searches published in
Refs.~\cite{PhysRevLett.106.131802,Aad2011186}, apparently without
encountering the case $\ul = 0$.  (A ``power constrained''
modification also used by ATLAS is described below.)

\begin{figure}
\centering
  \includegraphics*[width=1.0\textwidth,angle=90]{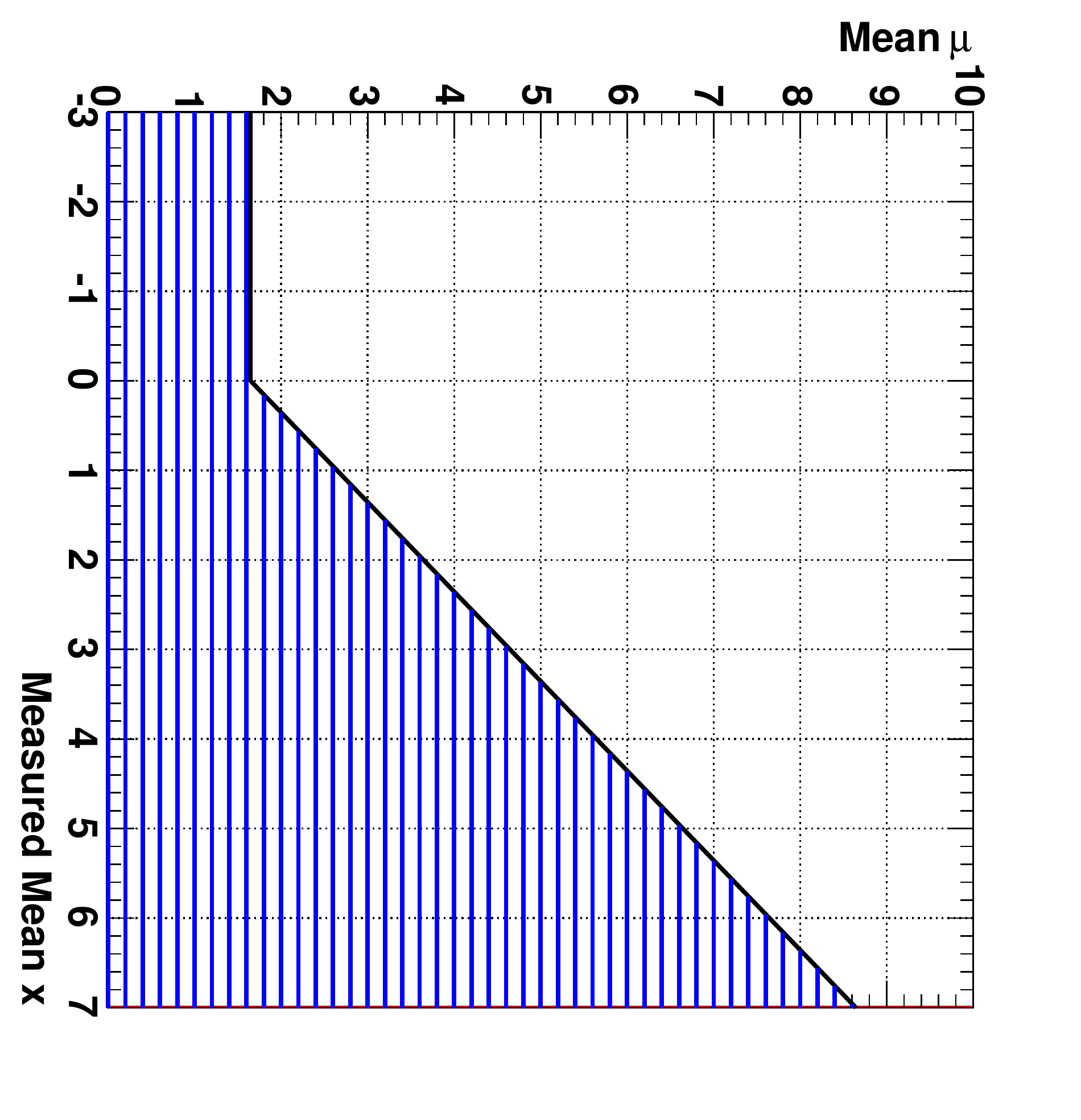}
\caption{Modification of the confidence belt in
Fig.~\ref{fig-uppershort} by using the $\ul$ for $x=0$ when $x<0$, so
that $\ul = \max(0,x) + 1.64$.  The same result is obtained by using a
Power Constrained Limit of CCGV \cite{PCL} with 50\% power
constraint. }
\label{fig-dhl}
\end{figure}

A more common notion was to use $\max(0,x)$ rather than $x$ in
Eqn.~\ref{ul95}, i.e., to move the measured value $x$ to the physical
boundary and proceed, obtaining $\ul = 1.64$.  The corresponding belt
(Fig.~\ref{fig-dhl}) has 100\% coverage for $\mu\le 1.64$.

\begin{figure}
\centering
  \includegraphics*[width=1.0\textwidth,angle=90]{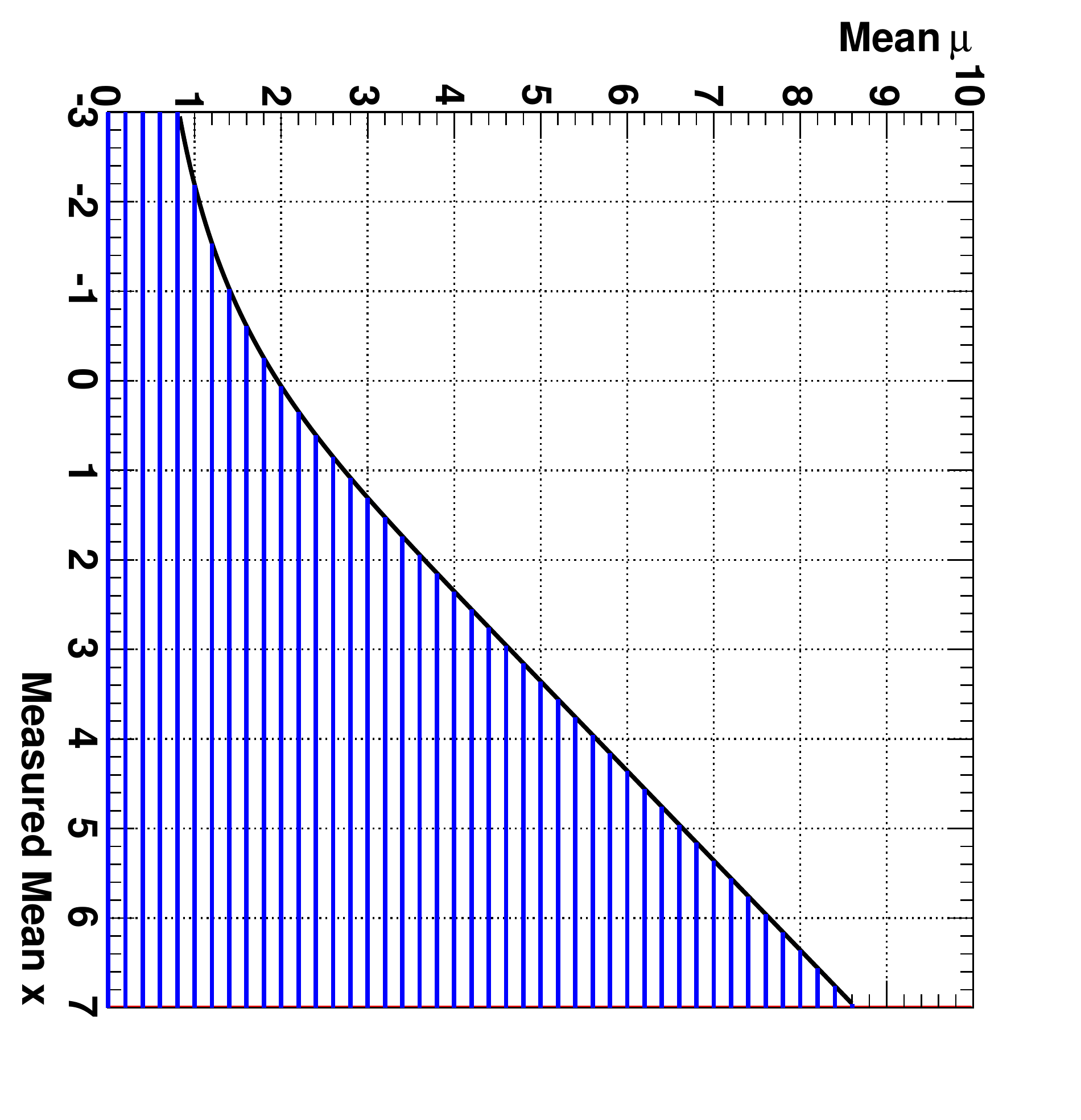}
\caption{Upper limits obtained via the Bayesian method recommended by
the PDG RPP, plotted as a confidence belt.  The prior probability
density for $\mu$ is uniform for all $\mu$ which exist in the model,
i.e., for $\mu\ge0$.  The horizontal lines contain more than 95\% of
the acceptance for $x$, so from the frequentist point of view the
upper limits are conservative. For this problem, the upper limits from
\cls\ are the same.}
\label{fig-bayes}
\end{figure}

\begin{figure}
\centering
  \includegraphics*[width=1.0\textwidth,angle=90]{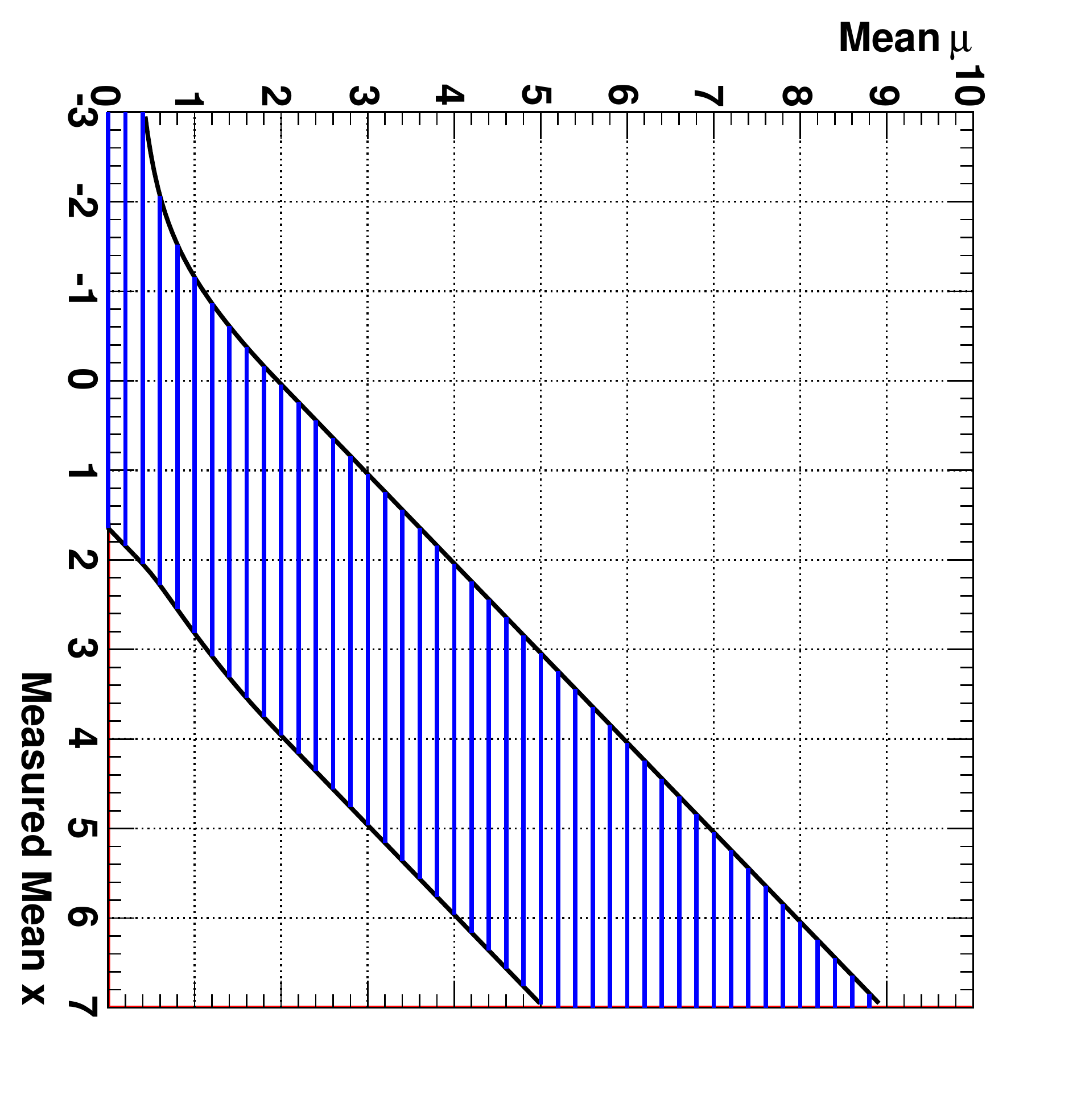}
\caption{95\% confidence belt advocated by Feldman and Cousins
\cite{feldman1998}.  For $x\le1.64$, the lower end of the interval is
0.  All horizontal acceptance intervals contain 95\% of the
probability for observing $x$.}
\label{fig-FCconstruct}
\end{figure}

There was however a sense among many (correct in my view) that the
problem with these diagonal line solutions with horizontal-line
modifications was fundamental and could not be patched up merely by
imposing a minimum value of $\ul$ or of $x$.  There was much
discussion in the 1980's and 1990's, leading to the 2000 Confidence
Limits Workshops \cite{clw,clwf} which evolved into the PhyStat
conference series.  During this period, three methods gained
significant support in HEP for constructing intervals that departed
from those based on the diagonal line: a Bayesian method (called
``very usual'' by Highland \cite{VH86} in 1986) with some basis in the
statistics literature; a method invented at LEP called \cls\
\cite{clw} that used reasoning apparently not in the statistics
literature and took advantage of some numerical coincidences; and a
method advocated by Feldman and Cousins~(F-C)~\cite{feldman1998},
which we learned was in Kendall and Stuart~\cite{kendall99}.  The
Particle Data Group's Review of Particle Physics (PDG RPP) \cite{PDG}
abandoned the diagonal line and described these three methods
beginning, respectively, in 1986, 2002, and 1998.  For this simple
problem, the Bayesian and CL$_{\rm S}$ belts are the same, shown in
Fig.~\ref{fig-bayes}.
The F-C belt, displayed in Fig.~\ref{fig-FCconstruct}, has a non-zero
lower edge for $x>1.64$, as discussed below.

\begin{figure}
\centering
  \includegraphics*[width=1.0\textwidth,angle=90]{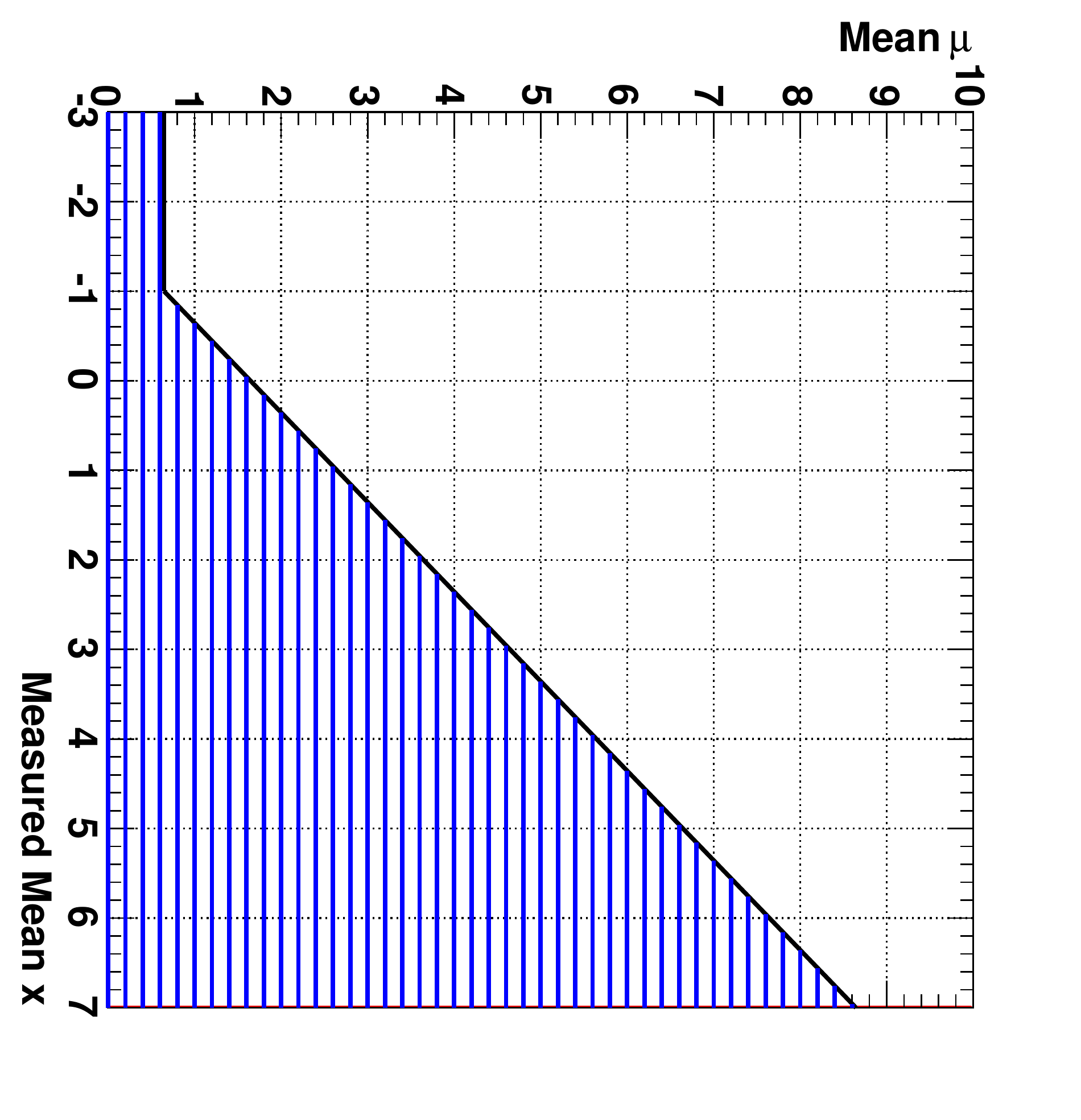}
\caption{Upper limits obtained by the ``Power Constrained Limit''
method following the recommended 16\% power constraint of CCGV
\cite{PCL} which leads to $\ul = \max(-1,x) + 1.64$.}
\label{fig-pcl}
\end{figure}

As an alternate protection against limits deemed to be ``anomalously
strong'' (such as those obtainable from Fig.~\ref{fig-trunc}), a
concept of ``Power Constrained Limit'' was advocated by Cowan,
Cranmer, Gross, and Vitells (CCGV) \cite{PCL}, and applied to ATLAS
Higgs searches \cite{atlashiggs,Aad:2011qi}, including the July 2011
submission in Ref.~\cite{atlashiggstautau}.  The method, as described
in some detail in Ref.~\cite{Aad:2011qi} and used in
Ref.~\cite{atlashiggstautau}, follows the recommendation of CCGV
\cite{PCL} (also cited by another ATLAS supersymmetry search
\cite{Aad2011398}), which for the example at hand yields $\ul =
\max(-1,x) + 1.64$.  (This corresponds to a power constraint
\cite{PCL} of 16\%.)  The corresponding belt is shown in
Fig.~\ref{fig-pcl}.  The old belt of Fig.~\ref{fig-dhl} turns out to
correspond to a Power Constrained Limit with a power constraint of
50\%, which was considered ``too extreme'' at the time Ref.~\cite{PCL}
was posted (16 May 2011).  Since then, PCL proponents and ATLAS have
reconsidered, and a more recent Higgs combination note
\cite{atlasconf112} uses a 50\% power constraint.

Thus it is timely to consider again what is problematic about the
original diagonal line of Eqn.~\ref{ul95}, and whether or not
modifications such as replacing $x$ with $\max(0,x)$ or $\max(-1,x)$
get to the root of these problems.

My views in 1998-2000, still largely unchanged, are discussed in the
Refs.~\cite{clw} and \cite{feldman1998}.  An issue that we called
``flip-flopping'' and its resolution is discussed in detail in
Ref.~\cite{feldman1998}. Another point in Ref.~\cite{feldman1998},
explained in more detail below in Sec.~\ref{sec-gof}, is how the
diagonal line of Fig.~\ref{fig-uppershort} undesirably couples
together goodness of fit for the model with interval estimation of
parameters of the model.  However, in most of this paper I focus on
following the additional leads given by statisticians in commenting on
a 2002 review paper by physicist Mark Mandelkern
\cite{mandelkern2002}.  In particular, the statistician Leon Jay
Gleser \cite{gleser2002} pointed to the literature on {\em
recognizable} and {\em relevant subsets} which gives great insight
into the problem from a different (though of course related)
perspective than we have had before.

The discussion points to quite remarkable theorems and examples in the
statistics literature.  The concept of ``most powerful hypothesis
test'' in the sense of Neyman and Pearson (N-P) can be in direct
conflict with a scientist's desire to extract the most relevant
inference from a particular data set at hand.  Desiderata have been
formulated in terms of pure conditional frequentist probabilities that
lead to a connection to Bayesian statistics, even though the
conditional frequentist probabilities discussed in this context have
the usual interpretation with the endpoints of the intervals as the
random variables.  The connection (still not completely known) has to
do with whether there exists {\em any} (possibly generalized) Bayesian
prior that leads to Bayesian credible intervals similar to the
confidence intervals in the frequentist confidence set; this can have
some relation to whether or not the sample space for $x$ has
recognizable subsets for which the experimenter {\em knows} that the
frequentist coverage is different from nominal in that subset.

If one's data is in such a recognizable subset, what to do is likely
context dependent.  But for searches for New Physics, I would side
with those who argue to make the frequentist-based inference as
relevant as it can be within the constraints of coverage.
(Ref.~\cite{feldman1998} still is my preferred way to do this.)  I
have also advocated for some time performing as well a Bayesian
analysis, which uses only the ``relevant'' probability of obtaining
the data at hand, but for which repeated sampling properties or prior
sensitivity may be unsatisfactory.  By comparing the two, one gets
even greater insight into any problem.

Section~\ref{simple} introduces Fisher's concept of recognizable
subsets of the sample space, using the example of the much-discussed
issue of empty intervals.  A reminder of the issue of coupling
goodness of fit and interval estimation is included at the end of this
section.  Section~\ref{buehler} describes the half-century-old
formalism and definitions for studying conditional coverage within
subsets, and some subsequent results and concepts from the statistics
literature.  Section~\ref{sec-xcrit} applies these concepts to the
original diagonal line of Eqn.~\ref{ul95}, thus revealing the
troublesome property, negatively biased relevant subsets, in the
language introduced in Sec.~\ref{buehler}.  Section~\ref{sec-power}
discusses methods which add a horizontal line to the original diagonal
line. Methods such as these, having 100\% unconditional coverage for
some values of $\mu$ while stating ``95\% C.L.'', are not considered
in the literature I have seen on relevant subsets; I believe that a
careful adaptation would raise some analogous concerns.  I conclude in
Sec.~\ref{concl}, along with comparisons to the intervals advocated by
F-C in Ref.~\cite{feldman1998}.

\section{A simple betting game}
\label{simple}
Suppose that Peter performs a set of repeated experiments, and after
each experiment, he uses the observed $x$ and Eqn.~\ref{ul95} to
announce, ``Using a procedure which is guaranteed to cover the unknown
true value of $\mu$ in 95\% of experiments, and fail to cover in 5\%
of experiments, I assert that the true value of $\mu$ is less than or
equal to $x+1.64$.''  Suppose then that Paula says, ``OK, in that
case, you should be willing to offer to bet against me at 19:1 odds
that each of your assertions is true. Let us play the following game.
After each experiment, I will decide, based on the value of $x$ you
obtain, whether or not to bet against the assertion you made following
that experiment, {\em and I will do this using no more information
about the model or $\mu$ than you have} (in particular without using
any prior knowledge of the true $\mu$).''  Peter says fine.

Paula then proceeds to bet against Peter's assertion whenever
$x<-1.64$, and to say ``no thanks'' to the bet whenever $x\ge -1.64$.
Paula not only wins in the long run -- in fact she wins every bet!  As
described in Sec.~\ref{sec-xcrit}, Paula can in fact win in the long
run by accepting all bets (at 19:1 odds) if $x<C$, where $C$ is any
constant of her choice. E.g., for $C=0$, she wins at least 10\% of the
bets, well above her break-even point of 5\%.

Since Paula is winning bets by using no more information than that
available to Peter, it is certainly arguable that Peter is not making
the most relevant assertions about each of his data sets.  This
example has much in common with a number of disturbing examples in the
statistics literature in which the N-P theory of tests that are most
powerful in the long run can lead to statements that appear to be
irrelevant or misleading for interpreting a particular data set at
hand.  The ``modern'' discussion seems to have been stimulated by Sir
Ronald Fisher (reprinted in \cite{fisher1990}) who coined the phrase
``recognizable subset'' in 1956 to describe a subset of ``entities''
for which it can be recognized that the probabilities associated with
entities in the subset are different from their (still purely
frequentist) probabilities in the superset of which the subset is a
part.

Classic examples include Sir David Cox's 1958 mixture experiment
\cite{cox1958} with two measuring devices with different $\sigma$, one
device chosen randomly as part of the repeated experimental procedure.
Another classic example is interval estimation for the mean $\mu$ of a
distribution uniform over $[\mu-1/2,\mu+1/2]$, based on a data set
consisting of two sampled values $x_1$ and $x_2$.  N-P procedures
based on power give the same confidence interval (or confidence limit)
for the data $\{x_1=0.99,x_2=1.01\}$ as for data
$\{x_1=.51,x_2=1.49\}$, even though the second set restricts $\mu$ to
the narrow range [0.99,1.01], while the first set only restricts $\mu$
to [0.51,1.49].  These two examples are particularly clean because in
each there exists an {\em ancillary statistic}, conceptually a
function of the data carrying information about the precision of the
measurement but no information about $\mu$.  (The ancillary statistic
is the index of the detector used in the Cox example, and $|x_2-x_1|$
in the second example.)  The ancillary statistics can be used to
divide the sample space into subspaces for calculating conditional
coverage probabilities relevant to the data at hand.

In these examples, there is a clear conflict between the criterion of
maximum power in N-P tests and the notion (however vague at this
point) of ``using all of the available relevant information in the
data set at hand''.  This conflict was pursued in a landmark 1959
paper by Robert Buehler \cite{buehler1959} (brought to our attention
by Gleser \cite{gleser2002}), in which he introduced a betting game
such as that above and defined the terminology as described in the
following section (which modernizes his Paul to Paula, but otherwise
mostly transcribes part of Buehler's paper). A key observation is that
even in the absence of ancillary statistics, one can sometimes place
bounds (away from the nominal C.L.!) on coverage probabilities within
recognizable subsets of the sample space.

\subsection{Coupling of goodness of fit and interval estimation}
\label{sec-gof}
Another, less complete, view \cite{feldman1998} of a difficulty of the
original diagonal line is that it couples together goodness of fit
(test of the model as a whole) with interval estimation (finding
preferred values of the parameters assuming that the model is true).
These two concepts are best kept separate, as is normally done in
curve-fitting when one uses the {\em magnitude} of $\chi^2$ at the
best-fit parameters as (only) a test of goodness of fit of the model,
while using $\Delta\chi^2$ with respect to the minimum value to obtain
an approximate confidence region for the parameters \cite{PDG}.  For
the Gaussian problem at hand of restricting $\mu$ based on measured
$x$, we have
\begin{equation}
\chi^2(\mu) = (x-\mu)^2;\ \mu\ge0.
\end{equation}

Let us consider the case where the measurement obtains $x=-1$. The
minimum $\chi^2$ is on the boundary, at $\mu=0$: $\chi^2_{\rm min} =
\chi^2(\mu=0) = 1.$ The upper limit from the diagonal line of
Eqn.~\ref{ul95} is $\ul = -1 + 1.64 = 0.64$.  We note that
$\chi^2(\mu=0.64) = (-1-0.64)^2 = 2.70$.  Thus, the 95\% upper limit
allows $\mu$ for which {\em absolute} $\chi^2\le2.70$.  But 2.70 is
the usual ``book value'' \cite{PDG} of the {\em difference}
$\Delta\chi^2$ to be used in computing a one-tailed upper limit!  The
fact that $\chi^2$ for $x=-1$ cannot be less than 1 for physical $\mu$
is somehow not used in computing the upper limit, as values with
$\Delta\chi^2 > 2.70 - 1 = 1.70$ are excluded.  This feature remains
with the Power Constrained Limit if the recommendation of CCGV
\cite{PCL} is followed (even though Ref.~\cite{PCL} uses
$\Delta\chi^2$), as a consequence of forcing the limit to be
one-sided.

In the same way, for $x<-1.64$, the {\em entire model} is rejected by
a goodness of fit test at 95\% C.L.  But if one accepts the model and
asks what values of the parameters are preferred {\em given that the
model is true}, it would seem somewhat useless to report the empty
set.

To repair the situation from this point of view, the F-C paper
\cite{feldman1998} advocates intervals based on $\Delta\chi^2$
considering values of $x$ on both sides of $\mu$ in the construction
of the acceptance intervals; for specified C.L., the critical value of
$\Delta\chi^2$ is associated with each $\mu$, not with $x$.  The
result is the belt in Fig.~\ref{fig-FCconstruct}.

\section{Buehler's betting game and subsequent literature}
\label{buehler}
Following Buehler \cite{buehler1959}, we let $A$ denote Peter's
assertion about $\mu$ being in a particular interval, and let $P(A)$
denote the (frequentist) probability that $A$ is true in the
unconditional sample space (all possible values of $x$).  If Peter's
intervals are confidence intervals obtained from a Neyman construction
with C.L. $\gamma$, then $P(A) = \gamma$, independent of $\mu$.  Using
knowledge of Peter's rule $R$ for constructing confidence intervals,
Paula adopts a strategy that consists of specifying in the
unconditional sample space (also called observation space) two subsets
$C^+$ and $C^-$ such that:
\begin{itemize} 
\item For observations in $C^+$, Paula bets that $A$ is true, risking
$\gamma$ to win $1-\gamma$.
\item For observations in $C^-$, Paula bets that $A$ is false, risking
$1-\gamma$ to win $\gamma$.
\end{itemize}
It is not required that a bet must always be made; i.e., the union of
$C^+$ and $C^-$ need not be the unconditional space.  To determine the
winner of each bet, we postulate the existence of a referee who knows
the true value of $\mu$.

We thus focus on the conditional probabilities $P(A|C)$.  Buehler
calls $P(A|C)-\gamma$ the {\em bias} of $C$. Then in typical interval
estimation problems, the bias of most subsets $C$ will not have the
same sign for all $\mu$.  Paula's task is to find subsets $C$ whose
bias has {\em the same sign for all $\mu$}.  These are called {\em
semi-relevant subsets induced by the rule $R$}.  If in addition the
bias is bounded away from zero, they are called {\em relevant}, a
stricter requirement than semi-relevant.  That is:
\begin{enumerate} 
\item  $C$ is a {\em semi-relevant subset} induced by the rule $R$ 
if either:
\begin{enumerate}
\item $P(A|C) > \gamma$ for all $\mu$, or
\item $P(A|C) < \gamma$ for all $\mu$; 
\end{enumerate}
\item $C$ is a {\em relevant subset} induced by the rule $R$ if, for
some constant $\const>0$ independent of $\mu$, either
\begin{enumerate}
\item $P(A|C) \ge \gamma+\const$ for all $\mu$, or
\item $P(A|C) \le \gamma - \const$ for all $\mu$;
\end{enumerate}
\end{enumerate}
where ``$>$'' or ``$\ge$'' indicates positive bias (overcoverage), 
and ``$<$'' or ``$\le$'' indicates negative bias (undercoverage).

Thus, for a negatively biased {\em semi-relevant} subset, there is
conditional undercoverage for all $\mu$, but there exist $\mu$ for
which the conditional undercoverage is arbitrarily small.  For a
negatively biased {\em relevant} subset, the conditional undercoverage
is bounded away from $\gamma$ by at least a finite amount $c$ for all
$\mu$.

All of these probabilities $P$ are {\em frequentist} probabilities:
assertions about confidence intervals/regions and limits are
assertions to be evaluated in terms of frequentist coverage in the
usual sense. That is, the endpoints of the interval (or boundary of
the confidence regions in higher dimensions) are the random variables,
not $\mu$, which is unknown and for which $P(\mu)$ need not be
defined.  The issue is about which ensemble to use for calculating
coverage properties for {\em post-data} inference, either the whole
sample space (as Neyman advocated), or some ``recognizable'' sub-space
$C$ in which the obtained data lies (as Fisher advocated).

One might hope that in a typical interval estimation problem, there
are no relevant or semi-relevant subsets, i.e., for any $C$, there
exists some $\mu$ for which $P(A|C) = \gamma$.  Buehler calls this
{\em strong exactness}; he refers to the usual unconditional coverage
$P(A) = \gamma$ as {\em weak exactness}.  But Buehler and subsequent
authors found semi-relevant subsets to be quite common in frequentist
confidence sets, so that ``nonexistence of semirelevant subsets is a
very severe requirement indeed.''  Relevant subsets have also been
identified in some famous problems, forcing one to think hard about
post-data inference in these cases.  A number of theorems were proven
giving necessary or sufficient conditions for the existence of
relevant or semi-relevant subsets.  Very significantly, a deep
relation with Bayesian theory was already noted by Buehler and by
David L. Wallace \cite{wallace1959} in the same year.  The ``very
severe requirement indeed'' of strong exactness was proven
\cite{wallace1959} to be automatically satisfied if there exists {\it
any} (proper) Bayesian prior such that each interval in the set of
frequentist confidence intervals also has Bayesian posterior
probability $\gamma$, i.e., each interval also has Bayesian
credibility $\gamma$.

The thread continued through key papers by Donald A. Pierce
\cite{pierce1973}, James V. Bondar \cite{bondar1977}, G.K. Robinson
\cite{robinson1979}, and very helpful reviews by George Casella
\cite{casella1992} and by Constantinos Goutis and George Casella
\cite{goutiscasella1995}.  Buehler's betting framework was
generalized, for example by letting Paula adjust the odds to make more
precise use of her conditional coverage calculation.  Pierce and
Robinson furthered the connections to Bayesian statistics,
generalizing the results showing that Bayesian procedures with proper
priors induce no semirelevant functions, and proving some more limited
statements about the converse.  (A succinct summary is in
Ref.~\cite{casella1992}.)  Connections were made to decision theory as
well.

Bondar referred to the absence of relevant or semi-relevant sets as
``consistency principles''.  By the time of his paper, there were
enough examples of otherwise-reasonable confidence sets admitting
semi-relevant subsets (of both signs of bias) and relevant subsets of
positive bias (i.e., overcoverage), that Robinson, Bondar, and others
seemed to reach a consensus that the criterion of ``elimination of
negatively biased relevant subsets was about right''
\cite{casella1992}.  This is about as much as one can demand within
the frequentist framework: to go further, one must use Bayesian
credible intervals and in some problems lose the guarantee of weak
exactness (unconditional coverage).  The complete connection between
conditional coverage and Bayesian procedures is still not known (in
particular for improper priors).

\section{Relevant subsets induced by the original diagonal line}
\label{sec-xcrit}
Remarkably, to my knowledge Gleser's Comment \cite{gleser2002} is the
first connection made in the statistics literature between all of this
relevant-subset theory and our HEP problem of a physically bounded
parameter:

\begin{quote}
The subset of samples having the property that the sample mean is two
standard deviations to the left of zero would have been called a
`recognizable subset' by Fisher (1956)\dots Professor Mandelkern's
example shows that the classic Neyman confidence interval is not
conditionally admissible in the case of estimating a positive mean.
Extension of this result to other cases of bounded parameters is
obvious.  In short, once something about the data is known, it is
possible for the frequentist properties of the confidence interval to
change: {\em the pre-data measure of risk is not necessarily the
correct post-data measure of uncertainty.}  ({}\cite{gleser2002},
italics in original.)
\end{quote}

Indeed it is not hard to work out the negative bias in relevant
subsets induced by the original diagonal line of Eqn.~\ref{ul95}. For
example, Paula can define a relevant subset $C$ in the sample space by
$x\,:\,x<0.7$.  So she bets against Peter's assertion $A$ at 19:1 odds
whenever $x<0.7$.  To see how she fares, we need to calculate, for
each $\mu$, the conditional coverage probability $P(\mu\le\ul\, |\,
x<0.7) = P(x \ge \mu - 1.64\, |\, x<0.7)$.  This probability is {\em
maximum} for $\mu=0$, in which case it is $P(x \ge -1.64\, |\,
x<0.7)$, for $x$ sampled from a Gaussian centered at 0.  That is
easily computed from two tails to be $1 - (0.05/0.758) = 93.4\%$,
negatively biased towards undercoverage. Paula concludes that the true
conditional odds in Peter's favor are {\em at most} 0.934/0.066, about
14:1, so she will win in the long run if Peter pays out at 19:1 odds
on the bets she makes.

\begin{figure}
\centering
  \includegraphics*[width=1.0\textwidth,angle=90]{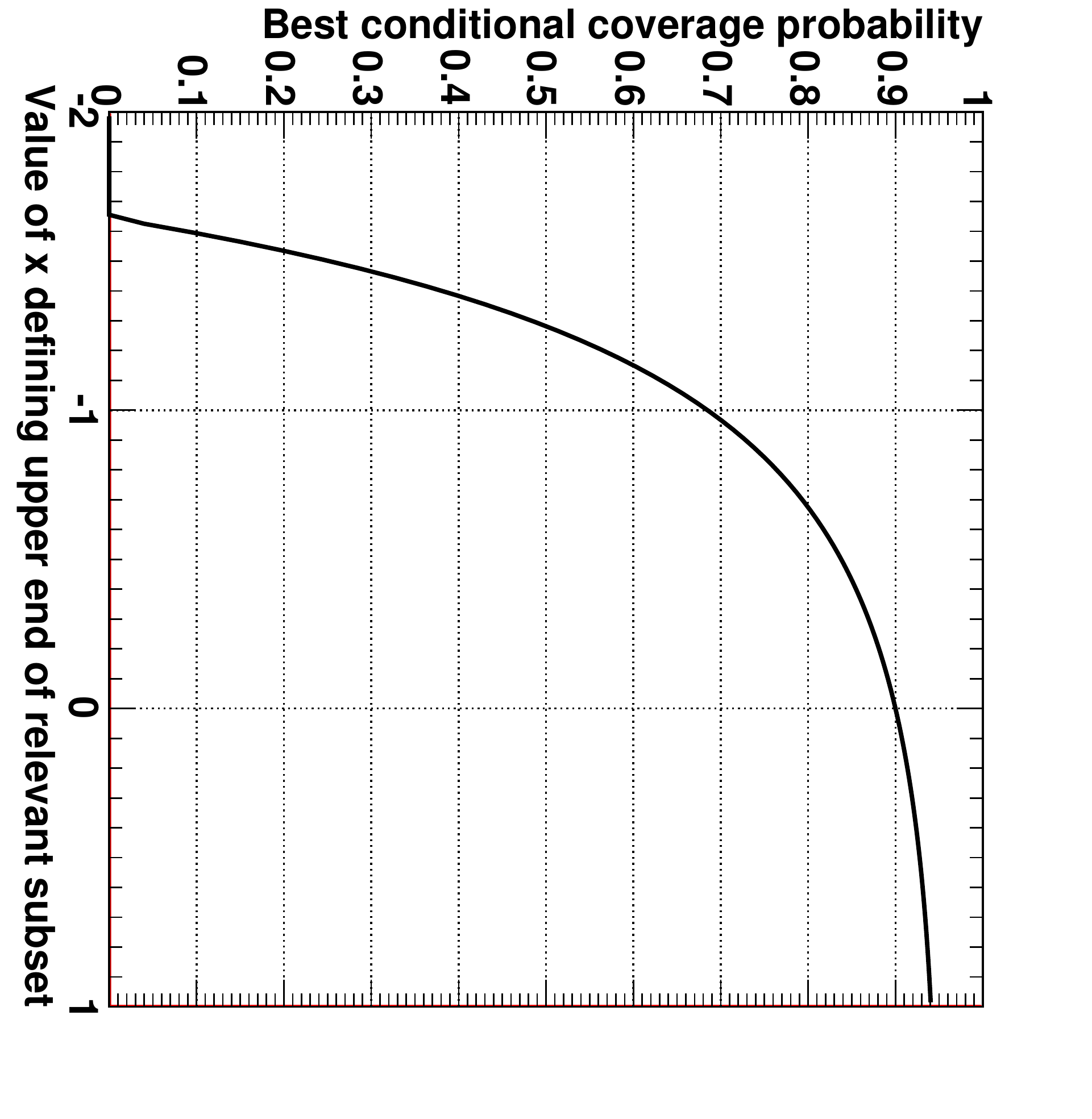}
\caption{Best conditional coverage $P(A|C)$ (for that $\mu$ having the
{\em highest} $P(A|C)$) vs.\ $\xcrit$, where Paula defines the
recognizable subset $C$ by $x<\xcrit$.  The recognizable subsets are
negatively biased relevant subsets since the conditional coverage is
less than of 0.95 by a finite amount.}
\label{fig-condcov}
\end{figure}

Figure~\ref{fig-condcov} displays the conditional coverage $P(A|C)$,
{\em for that $\mu$ having the highest $P(A|C)$}, where the relevant
subset $C$ is in the form of this example, defined by $x<\xcrit$, with
the value of $\xcrit$ on the horizontal axis.  As in the simple
betting example in Sec.~\ref{simple}, for $\xcrit<-1.64$, 0\% of
Peter's assertions are true. At $\xcrit=0$, the conditional coverage
increases to 90\% (with Type I error probability of 10\%, twice the
nominal 5\%).  For larger $\xcrit$, $P(A|C)$ further increases,
asymptotically approaching the unconditional 95\% as $\xcrit$ rises
above 1, and the effect of the boundary is less and less.

Figure~\ref{fig-condcov} thus quantifies for this problem the issue of
using only pre-data measures of uncertainty based on unconditional
probabilities (coverage, power) to distinguish among choices of
hypothesis tests.  I believe that it confirms physicists' intuition of
the past decades that the original diagonal line creates severe
problems for making relevant interpretations about $\mu$ from the data
set at hand.  What is perhaps new is that we can see readily that
these problems persist even beyond the ``obvious'' concerns that one
had when $x$ was less than $-1.64$ (empty-set or unphysical upper
limit), or when $x$ was only slightly larger than $-1.64$ (unnaturally
small upper limit with $\ul\ll \sigma$).

\section{\boldmath Confidence belts with 100\% acceptance intervals for some
values of $\mu$}
\label{sec-power}
The above theory of {\em relevant subsets} is based on the situation
in which the unconditional coverage of {\em all} values of $\mu$ is
that stated by the confidence level $\gamma$, 95\%.  This is the case
for the original diagonal line (Fig.~\ref{fig-uppershort}) and for the
F-C intervals (Fig.~\ref{fig-FCconstruct}).  As frequentist measures
do not average conditional coverage over $\mu$, the definitions are
all based on suprema of conditional coverage.  If there is any single
value of $\mu$ for which the horizontal acceptance interval for $x$ is
the {\em entire} horizontal line $(-\infty,+\infty)$, then the
coverage of the set of confidence intervals is 100\% for that value of
$\mu$.  This is the case, for example, with Fig.~\ref{fig-trunc},
which has 100\% coverage for $\mu=0$.  Thus all conditional coverage
calculations for $\mu=0$ also yield 100\%, and thus the supremum of
conditional coverage over all $\mu$ is also 100\%.  The {\em formal
theory} of relevant subsets is thus rendered moot by adding a single
value of $\mu$ with acceptance interval $(-\infty,+\infty)$ !  (A
100\% acceptance interval for $\mu = 3.14159$ would do so as well.)
Paula cannot be sure of winning bets from Peter in the long run,
because she will lose if the true $\mu$ is zero.

Should this evasion of the theory of relevant subsets, by having
acceptance intervals with 100\% acceptance, make physicists feel
better about the upper limits in Fig.~\ref{fig-trunc}?  I am not
inclined to set aside the insights of Sec.~\ref{sec-xcrit} simply
because the formal theory based on suprema is rendered moot.  The same
is true for methods which include yet more values of $\mu$ in the set
with 100\% coverage.  This fact remains: {\em If the true value of
$\mu$ is one for which the unconditional coverage is equal to the
stated confidence level of 95\%, then there exist sets $C$ for which
the conditional coverage of that $\mu$ is still bounded away from
95\%.}  Thus, the supremum of conditional coverage {\em over those
values of $\mu$ for which the unconditional coverage is 95\%} is also
bounded away from 95\%.

If we consider, for example, the Power Constrained Limit of
Fig.~\ref{fig-pcl} with $\ul = \max(-1,x) + 1.64$, Ref.~\cite{PCL}
points out that the coverage is 100\% for $\mu \le 0.64$, while
remaining exactly $\gamma = 95$\% for $\mu>0.64$.  We can thus
consider the conditional coverage properties for the entire set of
$\mu$ for which the unconditional coverage is 95\%.  I.e., in the
definitions in Sec.~\ref{buehler}, we interpret ``for all $\mu$'' to
mean ``for all $\mu$ for which the unconditional coverage is
$\gamma$''.  For any set $C$, we find the maximum conditional coverage
among the $\mu$ for which the unconditional coverage is $\gamma$.  We
observe that the conditional coverage of $\mu=0.65$ for $\xcrit=-1$ is
0, and the conditional coverage of $\mu=0.65$ for $\xcrit=0.65$ is
90\%.  That is, one shifts the curve in Fig.~\ref{fig-condcov}
horizontally by 0.64 to obtain the upper bound on the conditional
coverage among those $\mu$ for which the unconditional coverage is
stated (correctly) to be 95\%.

Is this a useful assessment of the conditional properties of Power
Constrained Limits?  Perhaps not, but I do not know a better way to
generalize the assessment of conditional properties of ``95\%
C.L. upper limits'' in the presence of acceptance intervals which have
100\% acceptance.  It seems to me that the burden should be on those
advocating PCL to explain how the theory of relevant subsets can be
adapted to this situation, so that it is possible to provide a more
useful critique.

\section{The frequentist alternative advocated by F-C}
\label{FC}

In the problem at hand, if one is willing to revisit the insistence on
one-sided limits, then there is a better performing alternative rule
$R$ for constructing confidence intervals, namely that advocated by
Gary Feldman and myself \cite{feldman1998}. This ``unified approach''
uses a likelihood-ratio (LR) ordering principle, i.e., inverting the
likelihood-ratio test described in Kendall and Stuart
\cite{kendall99}.  We set out to eliminate the empty-set intervals in
the much-discussed ``Simple Betting'' problem described in
Sec.~\ref{simple} in the context of searches for neutrino
oscillations, and discovered along the way that the problem was deeply
related to the insistence on having a one-sided limit, an insistence
that we believed was grounded more in convention rather than any deep
physics requirement.

If one wants to avoid empty-set intervals, then the logic leading from
exact coverage to the necessity of two-sided intervals (the ``unified
approach'' \cite{feldman1998} in which the lower endpoint of the
confidence interval is zero for only part of the sample space) is
quite straightforward.  In the Neyman construction, the acceptance
region for $\mu=0$ must include all $x\le0$ in order to guarantee no
empty-set intervals.  In order for the acceptance region for $\mu=0$
to contain exactly $\gamma = 95$\% of the sample space probability,
the upper endpoint must therefore be $x=1.64$.  Hence for $x>1.64$,
the confidence interval does not contain $\mu=0$.  This simple
argument is true as well for any other ordering rule (such as that
preferred by Mandelkern yielding a different set of unified intervals
\cite{mandelkern2002}) which has exact coverage $\gamma$ for all
values of $\mu$ and no empty-set intervals.

Incidentally, this ``one-tailed'' calculation of the upper endpoint of
the F-C acceptance region for $\mu=0$ also explains clearly why the
``unified approach'' delivers exactly what high energy physicists are
used to using in quantifying a discovery claim.  The classical
hypothesis test for rejecting $\mu=0$ at 5$\sigma$ is dual to
constructing confidence intervals at C.L. $\gamma =
1-2.8\times10^{-7}$ and checking if $\mu=0$ is in the confidence
interval. The F-C interval at this C.L. contains $\mu=0$ if and only
if $x\le 5\sigma$, as desired.

The unconditional coverage of the LR-ordering rule advocated by F-C is
exact by construction.  Since it is a frequentist rule which violates
the likelihood principle, perhaps there are semi-relevant subsets
induced by it, but I have not found them.  I would expect any
conditional bias induced by the LR-ordering to be vastly ameliorated
compared to that of the original diagonal line.

\section{\boldmath Should the inference about $\mu$ be independent 
of $x$ for $x<0$?}

Gleser's Comment \cite{gleser2002} on Mandelkern's review
\cite{mandelkern2002} made another deep point not yet mentioned in the
present paper.  If the model assumes that $\sigma$ is known exactly,
then the Likelihood Principle implies that one {\em should} make more
restrictive inferences about $\mu$ as $x$ becomes more negative; this
is the case for Bayesian upper limits (Fig.~\ref{fig-bayes}) and F-C
intervals (Fig.~\ref{fig-FCconstruct}), but not the case for the old
method of using $\max(0,x)$ (Fig.~\ref{fig-dhl}) or for the Power
Constrained Limit of Fig.~\ref{fig-pcl} (or for the version of the
unified approach advocated by Mandelkern).  Gleser notes,
\begin{quote}
\dots any confidence intervals that keep a constant width as $X$
becomes more negative, as some of the physicists seem to desire, are
indicating not necessarily what the data shows through the model and
likelihood, but rather desiderata imposed external to the statistical
model.
\end{quote}
As suggested by several Comments on Mandelkern's paper, if one is
unhappy with inferences becoming too restrictive, one should expand
the model to include uncertainty on $\sigma$.

\section{Discussion and conclusion}
\label{concl}
In retrospect, I believe that the HEP community that abandoned the
diagonal line of Eqn.~\ref{ul95} (for most of the issues of the PDG
RPP \cite{PDG} since 1987) understood a lot intuitively and from
studying many examples before and since.  With Gleser having pointed
us to ``relevant'' literature, we can now make the conditional
frequentist arguments which further illuminate the issues with
original diagonal line of Eqn.~\ref{ul95}.  The increased power that
one got from using the original diagonal line rather than the methods
in the PDG RPP seemed to incorporate inappropriate information (for
example goodness of fit to the model) that was hard to quantify.  Now
we can see that while the rule $R$ of the original diagonal line has
perfect coverage and maximal power against one-sided alternatives, it
induces severely negatively biased relevant subsets: one is making
assertions that under-cover (for all values of $\mu$) for data in the
recognizable relevant subset in which one's obtained measurement lies.

With a better understanding of the issue of conditional coverage, we
can also better understand why the Neyman-Pearson concept of pre-data
unconditional probabilities should not be trusted to address all these
difficulties.  Of course power is a tremendously useful concept which
we use in most contexts without conflict with other desiderata.  But
once the problem with the upper limits was identified as a conflict
between pre-data and post-data assessment of confidence, the
illuminating points naturally came from outside the Neyman-Pearson
paradigm, using ideas built on those of Neyman's great 20th century
frequentist rival, Sir Ronald Fisher.  Since the mathematics exposing
the bias induced by the original diagonal line is based on the
situation in which ``95\%'' really means ``95\%'', modifications to
include acceptance intervals of 100\% would seem to require more
generalized assessment tools.  I believe that the problem cannot be
easily dismissed in the absence of such tools: it is hard for me to
imagine that the underlying diseases of Fig.~\ref{fig-uppershort} are
eliminated simply by changing to Fig.~\ref{fig-trunc} with the
addition of a horizontal line at $\mu=0$.  As this is the step which
renders moot the relevant-subset literature, it is also not at all
clear that subsequently imposing a further step of ``power
constraint'' addresses this issue.

In the past, there had also been the rather vague notion that if there
was no Bayesian calculation (with {\em any} prior) that gave credible
intervals with some similarity to the confidence intervals, then the
frequentist calculation could be ``in trouble'' of some sort.  But it
was hard to quantify such ``violations of the Likelihood Principle'',
and these ideas were not always convincing to those claiming to be
``pure frequentist''.  Thus it is extremely enlightening to see the
theorems which relate Bayesian theory to the frequentist theory of
relevant subsets -- connections for which many of us in HEP had only
vague notions in the past.

If, in constructing confidence intervals/regions or limits, one has no
viable alternative but to use a particular rule $R$ that induces
severely negatively biased relevant subsets, then the situation would
seem to be quite unsatisfactory.  One might attempt to correct the
coverage assertion for bias, but problems with this were already noted
by Buehler: the sample space can have intersecting subsets having
biases that are different (and even of opposite sign).  A program of
research on conditional confidence including that of
Kiefer~\cite{kiefer1977} seems not to have converged in a general way.
There seems to be no general method for constructing confidence
intervals which is guaranteed to build in desirable coverage
properties in all recognizable subsets as well as in the superset.
(Seeking priors yielding Bayesian intervals with good coverage has
been suggested as a pragmatic approach.)  As a practical matter, one
is left to look for reasonable alternative rules $R$ that upon
inspection and in practice perform quite well in general.  In my
opinion, the two-sided LR-ordering rule advocated by F-C is such a
rule. The lower end of the interval is zero {\em unless} zero is
excluded in favor of non-zero values of $\mu$ by a {\em one}-tailed
test at C.L.$ = \gamma$; and at large $x$, the F-C interval naturally
approaches a two-sided central interval.

In conclusion, members of the community that developed the three
methods currently in the PDG RPP (Figs. \ref{fig-bayes} and
\ref{fig-FCconstruct}) were well aware of the possibility of
diagonal-line-based confidence belts such as those in
Figs.~\ref{fig-uppershort}, \ref{fig-trunc}, and \ref{fig-dhl}.  (I
do not know if anyone in that era ever advocated the belt in
Fig.~\ref{fig-pcl}).  It was quite reasonable that they fell out of
favor, in my opinion.  Taking into account insights accumulated since,
including those described in this note, I see no reason to return to
these or other variants of the diagonal line.

\medskip
{\bf Acknowledgments} I thank Luc Demortier, Tommaso Dorigo, Louis
Lyons, and Bill Murray for very helpful comments on earlier versions
of the manuscript, and Ofer Vitells and the other authors of
Ref.~\cite{PCL} for discussions about relevant subsets and Power
Constrained Limits. Of course, this does not imply their endorsement
of this note.  This work was partially supported by the U.S. Dept.\ of
Energy and by the National Science Foundation.


\bibliography{relevant}{}

\providecommand{\href}[2]{#2}\begingroup\raggedright\begin{thebibliography}{10%
}%
\makeatletter
\providecommand{\hrefCMSnoop }[0]{\@secondoftwo}%
\makeatother

\bibitem{VH86}
\hrefCMSnoop {} {V.~L. Highland, ``{Estimation of upper limits from
  experimental data}'',} Temple Univ., (1986).
\newblock COO-3539-38 (revised 1987, unpublished).

\bibitem{Aad:2009wy}
\hrefCMSnoop {} {{ ATLAS} Collaboration, ``{Expected Performance of the ATLAS
  Experiment - Detector, Trigger and Physics}'',}
  \href{http://www.arXiv.org/abs/0901.0512v4 [hep-ex]}{\texttt{
  arXiv:0901.0512v4 [hep-ex]}}. CERN-OPEN-2008-020. See pp. 1480 ff.

\bibitem{PhysRevLett.106.131802}
\hrefCMSnoop {} {{ ATLAS} Collaboration, ``{Search for Supersymmetry Using
  Final States with One Lepton, Jets, and Missing Transverse Momentum with the
  ATLAS Detector in $\sqrt{s}=7$ TeV $pp$ Collisions}'',} \textit{ Phys. Rev.
  Lett.} \textbf{ 106} (2011) 131802,
  \href{http://www.arXiv.org/abs/1102.2357v2 [hep-ex]}{\texttt{
  arXiv:1102.2357v2 [hep-ex]}}.
  \href{http://dx.doi.org/10.1103/PhysRevLett.106.131802}{\texttt{
  doi:10.1103/PhysRevLett.106.131802}}.

\bibitem{Aad2011186}
\hrefCMSnoop {} {{ ATLAS} Collaboration, ``Search for squarks and gluinos using
  final states with jets and missing transverse momentum with the {ATLAS}
  detector in proton-proton collisions'',} \textit{ Physics Letters B} \textbf{
  701} (2011) 186 -- 203, \href{http://www.arXiv.org/abs/1102.5290v1
  [hep-ex]}{\texttt{ arXiv:1102.5290v1 [hep-ex]}}.
  \href{http://dx.doi.org/10.1016/j.physletb.2011.05.061}{\texttt{
  doi:10.1016/j.physletb.2011.05.061}}.

\bibitem{PCL}
\hrefCMSnoop {} {{Glen Cowan, Kyle Cranmer, Eilam Gross, and Ofer Vitells},
  ``{Power-Constrained Limits}'',} \href{http://www.arXiv.org/abs/1105.3166v1
  [physics.data-an]}{\texttt{ arXiv:1105.3166v1 [physics.data-an]}}.

\bibitem{clw}
\hrefCMSnoop {} {F.~James, L.~Lyons, and Y.~Perrin, eds., ``Proceedings of the
  Workshop on Confidence Limits, CERN, Switzerland, 17-18 January 2000''}.
\newblock (2000).
\newblock The entire Proceedings is still interesting reading. CL$_{\rm S}$ is
  described in the paper by A. L. Read on p. 81.\\
  \url{http://cdsweb.cern.ch/record/411537/files/}.

\bibitem{clwf}
\hrefCMSnoop {} {{Louis Lyons}, ed., ``Workshop on Confidence Limits, Fermilab,
  Illinois, 27-28 March 2000''}.
\newblock (2000).
\newblock \\ \url{http://conferences.fnal.gov/cl2k/}.

\bibitem{feldman1998}
\hrefCMSnoop {} {G.~J. Feldman and R.~D. Cousins, ``{A Unified Approach to the
  Classical Statistical Analysis of Small Signals}'',} \textit{ Phys. Rev.}
  \textbf{ D57} (1998) 3873--3889,
  \href{http://www.arXiv.org/abs/physics/9711021}{\texttt{
  arXiv:physics/9711021}}. This paper also discusses in detail a second
  prototype problem and solution, that of Poisson counts in the presence of
  background. As emphasized by G\"unter Zech, violation of the Likelihood
  Principle is manifest when zero counts are observed.
  \href{http://dx.doi.org/10.1103/PhysRevD.57.3873}{\texttt{
  doi:10.1103/PhysRevD.57.3873}}.

\bibitem{kendall99}
A.~Stuart, K.~Ord, and S.~Arnold, ``Kendall's Advanced Theory of Statistics'',
  volume~2A.
\newblock Arnold, London, 6th edition, 1999.
\newblock and earlier editions by Kendall and Stuart. See the beginning page of
  the chapter on Likelihood Ratio Tests.

\bibitem{PDG}
\hrefCMSnoop {} {K.~Nakamura {et~al.}, ``{Review of Particle Physics}'',}
  \textit{ J. Phys.} \textbf{ G37} (2010) 075021. Current issue and archive is
  at the Particle Data Group's web site, {\\ \url{http://pdg.lbl.gov}}.
  \href{http://dx.doi.org/10.1088/0954-3899/37/7A/075021}{\texttt{
  doi:10.1088/0954-3899/37/7A/075021}}.

\bibitem{atlashiggs}
\hrefCMSnoop {} {{ ATLAS} Collaboration, ``{Higgs Boson Searches using the $H
  \rightarrow WW^{(*)} \rightarrow \ell\nu\ell\nu$ Decay Mode with the ATLAS
  Detector at 7 TeV}'',} ATLAS-CONF-2011-005, (Feb 11, 2011
  (revised Mar 10)).
\newblock The method including 16\% power constraint is described on p.\ 22. \\
  \url{http://cdsweb.cern.ch/record/1328619/files/ATLAS-CONF-2011-005.pdf}.

\bibitem{Aad:2011qi}
\hrefCMSnoop {} {{ ATLAS} Collaboration, ``{Limits on the production of the
  Standard Model Higgs Boson in $pp$ collisions at $\sqrt{s}$ =7 TeV with the
  ATLAS detector}'',} \href{http://www.arXiv.org/abs/1106.2748v2
  [hep-ex]}{\texttt{ arXiv:1106.2748v2 [hep-ex]}}.

\bibitem{atlashiggstautau}
\hrefCMSnoop {} {{ ATLAS} Collaboration, ``{Search for neutral MSSM Higgs
  bosons decaying to $\tau^+ \tau^-$ pairs in proton-proton collisions at
  $\sqrt{s}$ = 7 TeV with the ATLAS detector}'',}
  \href{http://www.arXiv.org/abs/1107.5003v1 [hep-ex]}{\texttt{
  arXiv:1107.5003v1 [hep-ex]}}. CERN-PH-EP-2011-104.

\bibitem{Aad2011398}
\hrefCMSnoop {} {{ ATLAS} Collaboration, ``Search for supersymmetry in $pp$
  collisions at $\sqrt{s}=7$ TeV in final states with missing transverse
  momentum and b-jets'',} \textit{ Physics Letters B} \textbf{ 701} (2011) 398
  -- 416, \href{http://www.arXiv.org/abs/1103.4344v1 [hep-ex]}{\texttt{
  arXiv:1103.4344v1 [hep-ex]}}.
  \href{http://dx.doi.org/10.1016/j.physletb.2011.06.015}{\texttt{
  doi:10.1016/j.physletb.2011.06.015}}.

\bibitem{atlasconf112}
\hrefCMSnoop {} {{ ATLAS} Collaboration, ``{Combined Standard Model Higgs Boson
  Searches in $pp$ Collisions at $\sqrt{s} = 7$ TeV with the ATLAS Experiment
  at the LHC}'',} ATLAS-CONF-2011-112, (July 24, 2011).
\newblock \\
  \url{http://cdsweb.cern.ch/record/1375549/files/ATLAS-CONF-2011-112.pdf}.

\bibitem{mandelkern2002}
\hrefCMSnoop {} {M.~Mandelkern, ``{Setting Confidence Intervals for Bounded
  Parameters}'',} \textit{ Statistical Science} \textbf{ 17} (2002), no.~2, pp.
  149--159. \\ \url{http://www.jstor.org/stable/3182816}.

\bibitem{gleser2002}
\hrefCMSnoop {} {L.~J. Gleser, ``{[Setting Confidence Intervals for Bounded
  Parameters]: Comment}'',} \textit{ Statistical Science} \textbf{ 17} (2002)
  pp. 161--163. \\ \url{http://www.jstor.org/stable/3182818}.

\bibitem{fisher1990}
R.~A. Fisher in \textit{ Statistical Methods, Experimental Design, and
  Scientific Inference: A Re-issue of Statistical Methods for Research Workers,
  The Design of Experiments, and Statistical Methods and Scientific Inference},
  J.~H. Bennett, ed.
\newblock Oxford University Press, Oxford, 1990.
\newblock See in particular pp. 113-114 of {\em Statistical Methods and
  Scientific Inference}.

\bibitem{cox1958}
\hrefCMSnoop {} {D.~R. Cox, ``{Some Problems Connected with Statistical
  Inference}'',} \textit{ The Annals of Mathematical Statistics} \textbf{ 29}
  (1958), no.~2, pp. 357--372. \\ \url{http://www.jstor.org/stable/2237334}.

\bibitem{buehler1959}
\hrefCMSnoop {} {R.~J. Buehler, ``{Some Validity Criteria for Statistical
  Inferences}'',} \textit{ The Annals of Mathematical Statistics} \textbf{ 30}
  (1959), no.~4, pp. 845--863. \\ \url{http://www.jstor.org/stable/2237430}.\\
  I follow some modern authors in using $\gamma$ for the C.L.; Buehler,
  following Neyman, uses $\alpha$ for the C.L., which in modern notation
  corresponds to $1-\alpha$.

\bibitem{wallace1959}
\hrefCMSnoop {} {D.~L. Wallace, ``{Conditional Confidence Level Properties}'',}
  \textit{ The Annals of Mathematical Statistics} \textbf{ 30} (1959), no.~4,
  pp. 864--876. \\ \url{http://www.jstor.org/stable/2237431}.

\bibitem{pierce1973}
\hrefCMSnoop {} {D.~A. Pierce, ``{On Some Difficulties in a Frequency Theory of
  Inference}'',} \textit{ The Annals of Statistics} \textbf{ 1} (1973), no.~2,
  pp. 241--250. \\ \url{http://www.jstor.org/stable/2958010}.

\bibitem{bondar1977}
\hrefCMSnoop {} {J.~V. Bondar, ``{A Conditional Confidence Principle}'',}
  \textit{ The Annals of Statistics} \textbf{ 5} (1977), no.~5, pp. 881--891.
  \\ \url{http://www.jstor.org/stable/2958515}.

\bibitem{robinson1979}
\hrefCMSnoop {} {G.~K. Robinson, ``{Conditional Properties of Statistical
  Procedures}'',} \textit{ The Annals of Statistics} \textbf{ 7} (1979), no.~4,
  pp. 742--755. \\ \url{http://www.jstor.org/stable/2958922}.

\bibitem{casella1992}
\hrefCMSnoop {} {G.~Casella, ``{Conditional Inference from Confidence Sets}'',}
  \textit{ Lecture Notes-Monograph Series} \textbf{ 17} (1992) pp. 1--12. \\
  \url{http://www.jstor.org/stable/4355622}.

\bibitem{goutiscasella1995}
\hrefCMSnoop {} {C.~Goutis and G.~Casella, ``{Frequentist Post-Data
  Inference}'',} \textit{ International Statistical Review / Revue
  Internationale de Statistique} \textbf{ 63} (1995), no.~3, pp. 325--344. \\
  \url{http://www.jstor.org/stable/1403483}.

\bibitem{kiefer1977}
\hrefCMSnoop {} {J.~Kiefer, ``{Conditional Confidence Statements and Confidence
  Estimators}'',} \textit{ Journal of the American Statistical Association}
  \textbf{ 72} (1977), no.~360, pp. 789--808. \\
  \url{http://www.jstor.org/stable/2286460}.

\end{thebibliography}\endgroup
\bibliographystyle{cms}
\end{document}